\documentclass[prb,twocolumn,showpacs,amsmath,amssymb,floatfix]{revtex4}
\usepackage{graphicx}
\begin{document}

\title{Ferromagnetic Properties of ZrZn$_2$}
\author{E. A. Yelland, S. J. C. Yates,  O. Taylor, A. Griffiths,  S. M. Hayden and A. Carrington}
\affiliation{H.~H. Wills Physics Laboratory, University of
Bristol, Tyndall Avenue, Bristol BS8 1TL, United Kingdom}

\date{\today}

\begin{abstract}
The low Curie temperature ($T_C \approx 28$\,K) and small ordered
moment ($M_0\approx0.17$\,$\mu_\mathrm{B}$f.u.$^{-1}$) of ZrZn$_2$
make it one of the few examples of a weak itinerant ferromagnet.
We report results of susceptibility, magnetization, resistivity
and specific heat measurements made on high-quality single
crystals of ZrZn$_{2}$. From magnetization scaling in the vicinity
of $T_C$ ($0.001<|T-T_C|/T_C<0.08$), we obtain the critical
exponents $\beta=0.52\pm0.05$ and $\delta=3.20\pm0.08$, and
$T_C=27.50\pm0.05$\,K. Low-temperature magnetization measurements
show that the easy axis is [111]. Resistivity measurements reveal
an anomaly at $T_C$ and a non-Fermi liquid temperature dependence
$\rho(T)=\rho_0+AT^{n}$, where $n=1.67\pm0.02$, for $1<T<14$\ K.
The specific heat measurements show a mean-field-like anomaly at
$T_C$. We compare our results to various theoretical models.
\end{abstract}



PACS, the Physics and Astronomy
\maketitle

\section{\label{sec:Intro} Introduction}

Ferromagnetism in the cubic Laves compound ZrZn$_2$ was discovered
by Matthias and Bozorth \cite{Matthias58} in 1958. Its occurrence
in ZrZn$_2$ is unusual because neither elemental Zr nor Zn is
magnetically ordered. The low Curie temperature and small ordered
moment of ZrZn$_2$ make it one of the few examples of a
small-moment or weak itinerant ferromagnet. ZrZn$_2$ was initially
considered to be a candidate for Stoner theory \cite{Stoner36}.
However a quantitative comparison of Stoner theory with experiment
suggests that spin fluctuation effects are important
\cite{Moriya1985Book}. In particular, the Curie temperature is
strongly renormalized downwards from the Stoner value estimated
from band structure parameters \cite{Kubler04}.

\begin{figure}
\begin{center}
\includegraphics[width=0.60\linewidth,clip]{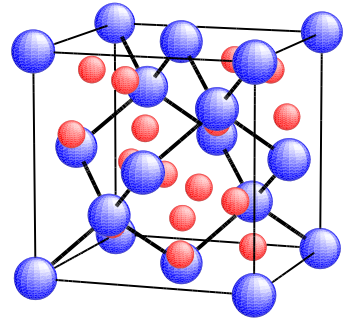}
\end{center}
\caption{ \label{fig:structure} (Color online) The C15 structure
of ZrZn$_2$. Zr atoms occupy the sites of the larger spheres.}
\end{figure}

ZrZn$_2$ crystallizes in the C15 cubic Laves structure shown in
Fig.1, with lattice constant $a=7.393$\,\AA.  The Zr atoms form a
tetrahedrally coordinated diamond structure and the magnetic
properties of the compound derive from the Zr 4d orbitals, which
have a significant direct overlap leading to the magnetic moment
being spread out over the network shown by the thick lines in
Fig.~\ref{fig:structure}. ZrZn$_2$ is strongly unsaturated: an
applied field of 5.7\,T results in a 50\,\% increase in the
ordered moment. In contrast, strong ferromagnets such as Fe and Ni
show a negligible increase of the ordered moment with field after
a single domain is formed. The unsaturated behavior of ZrZn$_2$
indicates a large longitudinal susceptibility and the presence of
longitudinal spin fluctuations \cite{Murata72,Lonzarich85}.
Further evidence for the existence of strong spin fluctuations in
ZrZn$_2$ is provided by the remarkably large effective mass of its
quasiparticles \cite{Yates03}. The low-temperature specific heat
coefficient \cite{Pfleiderer01c} and the calculated DOS at the
Fermi Level \cite{Yates03} imply an average mass enhancement $1 +
\lambda \approx 5$ at zero applied magnetic field
\cite{dHvAmassnote}. This is the largest known average mass
enhancement for a $d$-band metal, and is even slightly larger than
that of the strongly-correlated oxide system Sr$_2$RuO$_4$ for
which $1+\lambda \approx 3.6$
[Refs.~\onlinecite{MackenzieJDMRLMNF96,Bergemann03}].

It has been known for some time that the ferromagnetism of
ZrZn$_2$ is extremely sensitive to pressure \cite{Smith71}. Recent
experiments on the samples studied here \cite{Uhlarz04} have shown
that a pressure of $p_c=16.5$\,kbar causes the ferromagnetism to
disappear with a first order transition. Thus we may also view
ZrZn$_2$ as being close to a quantum critical point (QCP)
\cite{Uhlarz04,Kimura04}. In view of the strong longitudinal
fluctuations present in ZrZn$_2$ and its proximity to a QCP, it
has been proposed as a candidate for magnetically mediated
superconductivity \cite{Leggett78,EnzM78,FayA79,Pfleiderer01a}.
However, as discussed elsewhere \cite{Yelland05}, we find no
evidence for bulk superconductivity at ambient pressure in these
samples.

There are surprisingly few measurements of the fundamental
properties of high-quality samples (RRR$\geq$100) of ZrZn$_2$ in
the literature. In this paper, we present a study of the transport
and thermodynamic properties of ZrZn$_2$. In particular, we have
measured the magnetocrystalline anisotropy, the magnetization
isotherms for $2<T<40$\,K and $0<B<5$\,T; the resistivity from low
temperatures through the Curie temperature; and the specific heat
capacity for $0.3<T<40$\,K. These properties contain information
about the quasiparticles and magnetic interactions in ZrZn$_2$.
For example, the temperature dependence of the resistivity at low
temperatures gives information about the fundamental excitations
that scatter quasiparticles, and the comparison of specific heat
and magnetization data gives information about the importance of
spin fluctuations in ZrZn$_2$.

\section{\label{sec:Exp} Experimental Details}

\subsection{\label{sec:growth}Crystal Growth and Sample Quality}

ZrZn$_2$ melts congruently at 1180$^\circ$C
[Refs.~\onlinecite{Elliot65,Massalski86}]. At this temperature
zinc has a vapor pressure of about 10~bars and is an aggressive
flux. Thus we chose to grow ZrZn$_2$ by a directional cooling
technique \cite{Schreurs89}. Stoichiometric quantities of
high-purity zone-refined Zr (99.99\%, Materials Research MARZ
grade) and Zn (99.9999\%, Metal Crystals) were loaded into a
Y$_2$O$_3$ crucible. The total charge used was 4.2~g. The crucible
was sealed inside a tantalum bomb which was closed by electron
beam welding under vacuum.  The assembly was heated to
1210$^\circ$C and then cooled through the melting point at
2$^\circ$C hr$^{-1}$.  The ingot was then annealed by cooling to
500$^\circ$C over a period of 72\,hr. This method produced, on
occasions, single crystals of volumes up to approximately
0.4\,cm$^{3}$. Single crystals produced in this way had residual
resistivities as low as $\rho_0$= 0.53\,$\mu \Omega \mathrm{cm}$
corresponding to a residual resistance ratio RRR~=
$\rho(293\,\mathrm{K})/\rho(T \rightarrow 0)$=105. With the
exception of Ref.~\onlinecite{vanDeursen86}, previous reports by
other groups of the fundamental transport and thermodynamic
properties of ZrZn$_2$ have been carried out on samples with
RRR\,$\lesssim$\,45.

The residual resistivity $\rho_0$ and the Dingle temperature
determined from de Haas-van Alphen measurements may be used to
estimate the quasiparticle mean free path $\ell$ due to impurity
scattering. For a crystal with cubic symmetry \cite{ZimanBook},
\begin{equation}
\sigma = \frac{1}{4 \pi^{3}} \frac{e^{2} \tau}{\hbar} \frac{1}{3}
\int v_{F} \; dS_{F}
= \frac{1}{4 \pi^{3}} \frac{e^{2}}{\hbar} \frac{1}{3}
\int \ell \; dS_{F}.
\end{equation}
From band structure calculations \cite{Yates03}, we estimate the
sum of the Fermi surface areas to be $S_{F}=$\ $1.9 \times 10^{21}
\mathrm{m}^{-2}$. Hence $\ell_{\mathrm{trans}}=1350$\,\AA. A
second estimate of the quasiparticle mean free path can be made
from the de Haas-van Alphen effect \cite{Yates03}; values are in
the range $\ell_{\mathrm{dHvA}} =$\ 1500--2800\,\AA\ depending on
Fermi surface orbit, in approximate agreement with
$\ell_\mathrm{trans}$. In general, one expects that
$\tau_{\mathrm{dHvA}} < \tau_{\mathrm{trans}}$ since
$\tau_{\mathrm{trans}}$ is weighted towards large momentum changes
\cite{Abrikosov63}, whereas $\tau_{\mathrm{dHvA}}$ weights all
scattering equally. However, the opposite situation
$\tau_{\mathrm{dHvA}} > \tau_{\mathrm{trans}}$ may also arise in
an inhomogeneous sample, since the exponential scattering rate
dependence of the Dingle factor $R_{D}=\exp(- \pi m_{b} / e B
\tau)$ causes $\ell_\mathrm{dHvA}$ to be strongly weighted towards
high quality regions of the sample. Given these considerations,
the two mean free path estimates $\ell_\mathrm{dHvA}$ and
$\ell_\mathrm{trans}$ are as consistent as can be expected.

\subsection{\label{sec:measure_trans}Transport and Thermodynamic Measurements}
Resistivity measurements were made using a standard a.c.\
technique using a Brookdeal 9433 low-noise transformer and SR850
digital lock-in amplifier with a measuring frequency $f$=2~Hz.
Sample contacts were made with Dupont 4929 conducting Ag/epoxy.
Measurements of a.c.\ susceptibility were made by a standard
technique in which the sample was mounted inside a small coil of
approximately 2500 turns.  The system was calibrated using the
superconducting transition of an indium sample of similar size to
the ZrZn$_2$ sample. Magnetization measurements were made using a
commercial Quantum Design MPMS-XL SQUID magnetometer.

Heat capacity measurements were made both by a long-pulse method
and an a.c.\ method \cite{Carrington97}. In the long-pulse
technique \cite{ShepherdJP1985} the sample was mounted on a
silicon platform connected to a temperature-controlled stage by a
thin copper wire. In the a.c.\ technique the sample was mounted on
a flattened 12\,$\mu$m chromel-constantan thermocouple and heated
optically. The long pulse technique allows an accurate
determination of the absolute value of $C(T)$ but has relatively
low resolution. The a.c.\ technique has much higher resolution
($\sim0.1$\%), and is thus ideally suited to resolving the small
anomaly at $T_C$. However, it does not allow an accurate
determination of the absolute value of $C(T)$ due to the poorly
defined addenda contribution. We estimated the addenda by
measuring a Cu sample with similar $C(T)$; the a.c.\ data with the
estimated addenda subtracted were multiplied by a scale factor to
match the long-pulse data at a single temperature.

\section{Results}

\subsection{Magnetization Isotherms and Scaling near the Ferromagnetic Transition}

\subsubsection{Low Temperature Isotherms}

\begin{figure}
\begin{center}
\includegraphics[width=0.95\linewidth,clip]{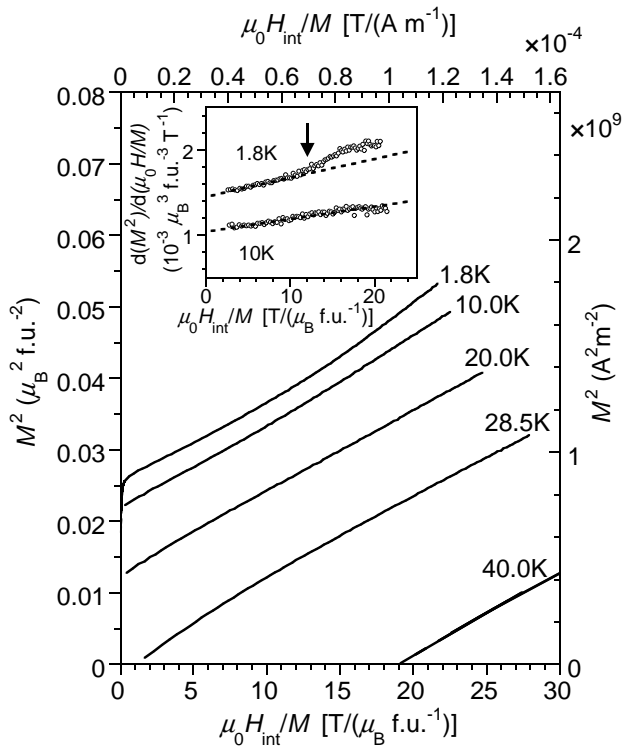}
\end{center}
\caption{ \label{fig:zrzn2_arrott} Arrott plot of magnetization
isotherms in the temperature range $1.8\leq T \leq 40$\,K. In the
Stoner theory of a weak itinerant ferromagnet, $M^2$ is a linear
function of $H/M$. Inset: the derivative
$\mathrm{d}(M^2)/\mathrm{d}(\mu_0 H/M)$ of the 1.8\,K and 10\,K
Arrott curves; the 1.8\,K curve is offset by +5$\times
10^{-3}\,\mu_\mathrm{B}^{3}\,\mathrm{f.u.}^{-3}\,\mathrm{T}^{-1}$.
Dotted lines are guides to the eye. Note the feature in the 1.8\,K
curve, indicated by an arrow.}
\end{figure}
Since the discovery of ZrZn$_2$ there have been many studies of
the magnetic properties. Previous work
\cite{Matthias58,OgawaS67,Knapp71,Smith71,Ogawa76,Mattocks1978,
vanDeursen86,Grosche95,Seeger95,Pfleiderer01a,UhlarzPH04} has
shown that the Curie temperature and the ordered moment, are
strongly dependent on sample quality and composition
\cite{Knapp71}. Thus the purpose of the present magnetization
measurements is to attempt to characterize the magnetic properties
in the clean limit. The high-quality of our samples is indicated
by their low residual resistivity (RRR=105) and long mean free
path, which has allowed much of the Fermi surface to be observed
by the dHvA effect \cite{Yates03}.

Although many plots of the magnetization isotherms for ZrZn$_2$
have been reported
\cite{OgawaS67,Knapp71,Mattocks1978,vanDeursen86} improved
instrumentation and higher sample quality have recently allowed
fine structure to be observed. For a weak ferromagnet, the
magnetization isotherms are generally expected to obey the Stoner
form in which $M^2$ is a linear function of $H/M$.
Fig.~\ref{fig:zrzn2_arrott} is an Arrott plot of our magnetization
isotherms. The near linearity of the isotherms confirms that they
are indeed of approximately the expected form. However, the
isotherm at $T=1.8$\,K shows an intriguing feature at $\mu_0
H\approx 2.4$\,T, indicated by an arrow in
Fig.~\ref{fig:zrzn2_arrott}. Another more pronounced feature is
directly visible in the $T=1.75$\,K $M(H)$ isotherm plotted in
Ref.~\onlinecite{Pfleiderer01a}, at a field $\approx 6$\,T,
outside the range of our present measurements. Structure in the
electronic DOS close to the Fermi level is expected to have a
profound influence on $M(T,B)$ in itinerant ferromagnets
\cite{Shimizu64,Shimizu65}, and calculations \cite{Yates03}
suggest that just such structure is present in the DOS of
ZrZn$_2$, the Fermi level lying between two sharp peaks in the
majority spin DOS. This is an interesting connection that could be
explored in future work.

\subsubsection{Scaling near $T_C$}

\begin{figure}
\begin{center}
\includegraphics[width=0.95\linewidth,clip]{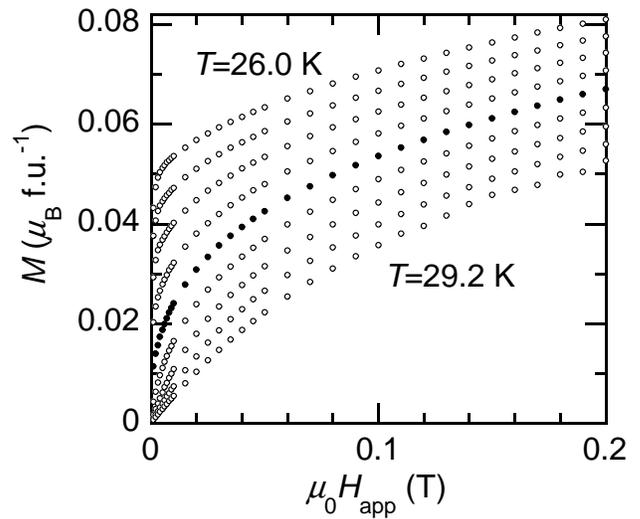}
\end{center}
\caption{ \label{fig:zrzn2_critical_isotherms} Low field
magnetization isotherms near the ferromagnetic critical
temperature of ZrZn$_2$ (near-critical isotherm at $T=27.6$\,K
shown by filled circles).}
\end{figure}

\begin{figure}
\begin{center}
\includegraphics[width=0.95\linewidth,clip]{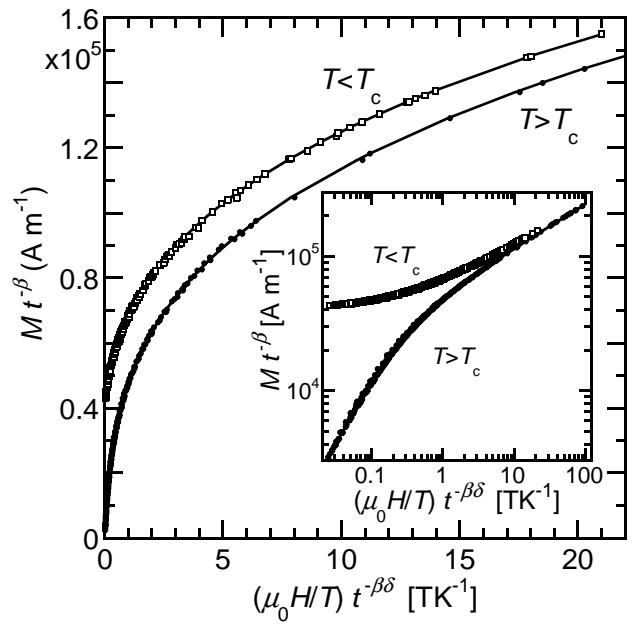}
\end{center}
\caption{ \label{fig:zrzn2_scaling} Scaling plot of magnetization
isotherms of ZrZn$_2$ near $T_C$. The data shown here are the same
as those in Fig.~\ref{fig:zrzn2_critical_isotherms} except that we
have excluded the region $\mu_0 H<0.005$\,T to avoid systematic
errors due to multiple ferromagnetic domains inside the sample
(see text). The correct choice of $T_C$, $\beta$ and $\delta$
causes the scaled data to collapse onto a single curve for $T<T_C$
and another for $T>T_C$. The values that allow the best fit to the
scaled data of a 7$^\mathrm{th}$ order polynomial in $M^*$ are
shown in Table~\ref{table:exponents}. The solid lines show the
best fit polynomials for $T<T_C$ and $T>T_C$. The inset shows the
same data on log-log axes.}
\end{figure}
At temperatures approaching the Curie temperature of a ferromagnet
the magnetization isotherms $M(H)$ become highly nonlinear making
the identification of $T_C$ difficult. An accurate determination
of $T_C$ can be achieved by a scaling analysis of the
magnetization close to the transition; we have therefore measured
magnetization isotherms at a set of temperatures near $T_C$
allowing us to determine $T_C$ and the magnetization scaling
exponents.

Close to the critical temperature of a ferromagnetic transition,
the magnetization $M(H,T)$ can be expressed as $M^*(h^*)$ where
$M^*=M t^{-\beta}$ and $h^*=(H/T) t^{-\beta\delta}$ are
appropriately scaled quantities \cite{YeomansBook}. Here
$t=|1-T/T_C|$ is the reduced temperature and $\beta$, $\delta$ are
the critical exponents. The experimental determination of $\beta$
and $\delta$ allows remarkably universal conclusions to be drawn
about the physical model which underlies the phase transition. We
have measured the magnetization $M(H)$ in applied fields $0<\mu_0
H<0.2$\,T at closely spaced temperatures in the range $0.1\leq
|T-T_C|<2.2$\,K. Fig.~\ref{fig:zrzn2_critical_isotherms} shows the
raw data. In order to determine $T_C$ and the scaling exponents,
$M^*$ and $h^*$ were calculated from the experimental data set for
all values of $\beta, \delta$ and $T_C$ within a certain volume of
the 3D parameter space encompassing both the Heisenberg and mean
field models. For each combination of values, the polynomial in
$M^*$ which best fit the scaled experimental values of $h^*$ was
found, and its goodness-of-fit ($\chi^2$) value calculated. The
correct parameter values are taken as those for which the best fit
polynomial has the lowest $\chi^2$. Only data points for which
$\mu_0 H_\mathrm{app}\geq0.005$\,T are included in the scaling
analysis in order to ensure the formation of a single
ferromagnetic domain inside the sample. The data have been
corrected for the small demagnetizing factor of this sample,
$D=0.060$, but we note that the final results are barely affected
by this. We find $T_C=(27.50\pm0.05)$\,K, $\beta=0.52\pm0.05$ and
$\delta=3.20\pm0.08$ (see Table~\ref{table:exponents}). The values
of $\beta$ and $\delta$ are close to those obtained on lower
quality samples with smaller $T_C$'s \cite{Seeger95}.

In order to interpret our scaling results, we need to know whether
we are in the critical region. The Ginzburg criterion
\cite{ChaikinBook},
\begin{equation}
\Delta T_\mathrm{G}=\frac{T_C k_\mathrm{B}^2}{32\pi^2(\Delta
C_\mathrm{V})^2\xi_0^6}
\end{equation}
allows us to estimate the extent of the critical region. From
neutron scattering measurements \cite{Bernhoeft88} of the
wavevector-dependent magnetic susceptibility $\chi(q)$ and low
temperature magnetization measurements we estimate the magnetic
correlation length $\xi_0=33$\,\AA. The specific heat jump is
$\Delta C\approx155$\,mJ\,K$^{-1}$\,mol$^{-1}$ (see
section~\ref{sec:spec_heat}). Hence $\Delta T_\mathrm{G}=0.4$\,mK.
Thus our data are collected outside the critical region where mean
field behavior is expected. Table~\ref{table:exponents} shows that
our results are indeed in agreement with the expected mean field
results for a 3D ferromagnet. Finally, is it worth commenting that
determining the Curie temperature $T_c$ from critical scaling, as
described here, yields a lower value than that obtained by taking
the peak in $(dM/dB)$ with $T$.  This may explain the slightly
higher values reported elsewhere \cite{Pfleiderer01a} for similar
samples.

\begin{figure}
\begin{center}
\includegraphics[width=0.95\linewidth,clip]{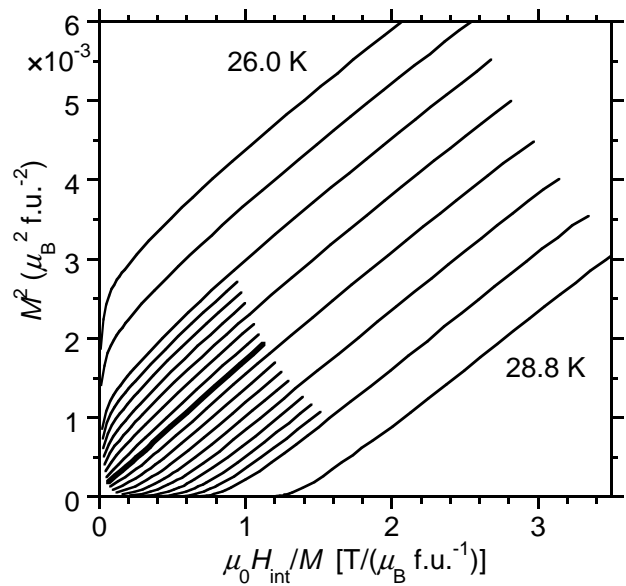}
\end{center}
\caption{ \label{fig:zrzn2_arrott_Tc} Arrott plot of magnetization
isotherms near $T_C$ for ZrZn$_2$. The isotherms were measured at
0.4\,K intervals and at 0.1\,K intervals close to $T_C$; the bold
line is the critical isotherm at $T=27.5$\,K. In the Stoner theory
of a weak itinerant ferromagnet, $M^2$ is a linear function of
$H/M$}
\end{figure}

\begin{table}
\caption{\label{table:exponents} Theoretical critical exponents
for various models of phase transition and the experimental values
found in this work. $T_C$ was found to be 27.50$\pm$0.05\,K.}
\begin{ruledtabular}
\begin{tabular}{c|ccc}
  & mean-field  &  Heisenberg & this work  \\
\colrule
$\beta$ & 0.5 & 0.326 &  0.52 $\pm$ 0.05 \\
$\delta$ & 3 & 4.78   &  3.20 $\pm$ 0.08 \\
\end{tabular}
\end{ruledtabular}
\end{table}

\begin{figure}
\begin{center}
\includegraphics[width=0.95\linewidth,clip]{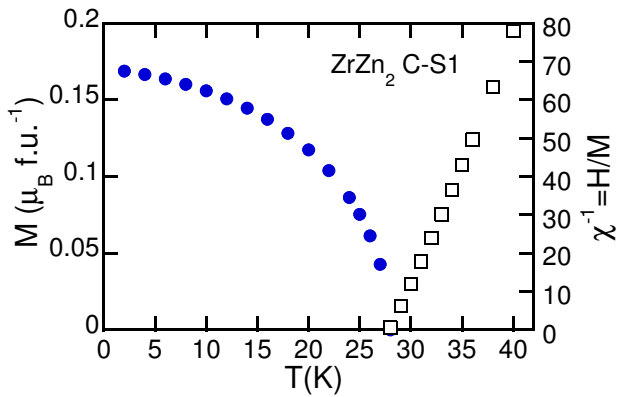}
\end{center}
\caption{ \label{f:zrzn2_M_chi T} The temperature dependence of
the magnetization (filled circles) and the inverse susceptibility
(open squares) as determined from Arrott plots.}
\end{figure}

\subsection{Magnetocrystalline Anisotropy}

\begin{figure}
\begin{center}
\includegraphics[width=0.95\linewidth,clip]{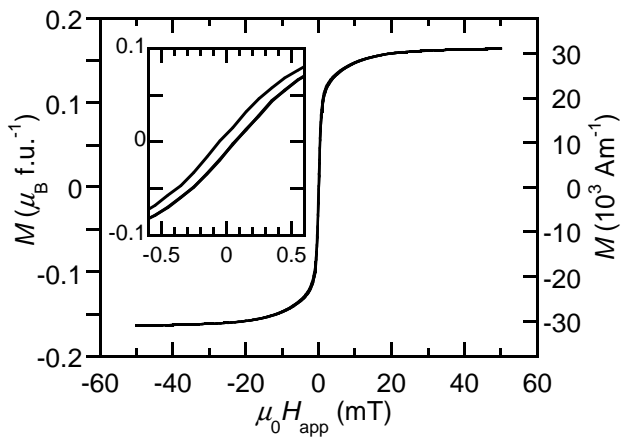}
\end{center}
\caption{ \label{f:zrzn2_hyst} Magnetic hysteresis curve for
ZrZn$_2$ measured at $T=5$\,K. The inset shows the central region
magnified. Note the small coercive field $H_\mathrm{coer}\approx
0.05$~mT.}
\end{figure}

The exchange interaction in an itinerant electron system is often
modelled as isotropic, depending only on the relative orientation
of electron spins. However, the crystal field also enters the free
energy via spin-orbit coupling, causing anisotropy of magnetic
properties. One motivation for studying the magnetocrystalline
anisotropy is that it has implications for the symmetry of the
superconducting order parameter in the spontaneous mixed state of
the putative superconducting phase of ZrZn$_2$
[Ref.~\onlinecite{Walker02}].

Fig.~\ref{f:zrzn2_hyst} shows a low-field hysteresis loop measured
for applied fields in the range $-0.05$\,T$<\mu_0
H_\mathrm{app}<0.05$\,T. The main panel shows that a field of
approximately 5\,mT is sufficient to create a single ferromagnetic
domain. The inset reveals the low coercive field of the present
samples $H_\mathrm{coer}\approx0.05$\,mT and justifies our method
(see below) for obtaining the magnetic anisotropy constants.

\begin{figure}
\begin{center}
\includegraphics[width=0.95\linewidth,clip]{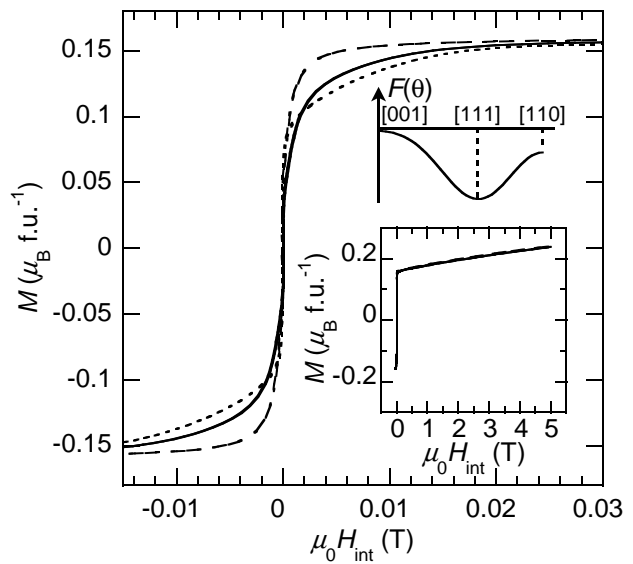}
\end{center}
\caption{ \label{fig:magnetization_anisotropy} Magnetization
curves of a ZrZn$_2$ single crystal measured at $T=5$\,K with field
applied along [100] (dotted line), [110] (solid line) and [111]
(dashed line). The ferromagnetic easy axis is seen to be
[111] at this temperature. Upper inset: plot of Eq.(2) using
values of $K_1$ and $K_2$ determined from
$M_\parallel(H_\mathrm{int})$ as described in the text. Lower
inset: $M_\parallel(H_\mathrm{int})$ up to 5\,T for the same
sample.}
\end{figure}

We have determined the ferromagnetic anisotropy of ZrZn$_2$ from
$M(H)$ isotherms at $T=5$\,K. A disc-shaped sample (diameter
$2R=2.72$\,mm, thickness $t=0.47$\,mm) was spark-cut so that the
plane of the disc was (110). This geometry allows access to the
three major cubic symmetry directions [100], [110] and [111] with
the magnetic field in the plane of the disc so that the
demagnetizing field is the same in each orientation. For this
sample, the volume averaged demagnetizing factor is
$D\approx0.146$\ [Ref.~\onlinecite{Crabtree77}], which for
$M=0.16\,\mu_\mathrm{B}\,\mathrm{f.u.}^{-1}$ corresponds to an
average demagnetizing field $\mu_0 H_\mathrm{D}\approx0.037$\,T
inside the sample. In this section we denote the measured
magnetization by $M_\parallel$ to emphasize that the SQUID
measurement is only sensitive to the component of $\mathbf{M}$
parallel to $\mathbf{H}_\mathrm{app}$, which is vital for
extracting the anisotropy constants by the thermodynamic method
described later. Fig.~\ref{fig:magnetization_anisotropy} shows
$M_\parallel(H_\mathrm{int})$ measured with $H_\mathrm{app}$
parallel to each of the three symmetry directions;
$H_\mathrm{int}$ is the internal field that, in a simple scalar
notation, is approximately related to the applied field
$H_\mathrm{app}$ by $H_\mathrm{int}=H_\mathrm{app}-DM$. As
$H_\mathrm{int}$ increases from zero, a single domain is rapidly
formed. The slower increase in $M_\parallel(H_\mathrm{int})$ for
$\mu_0 H_\mathrm{int}> 0.003$\,T with $\mathbf{H}_\mathrm{app}
\parallel [001]$ or $[110]$, arises from the rotation of $\mathbf{M}$ away from the easy axis towards
$\mathbf{H}_\mathrm{int}$ as the ratio of the interaction term
$\mathbf{M} \cdot \mathbf{H}_\mathrm{int}$ to the anisotropy
energy increases. Fig.~\ref{fig:magnetization_anisotropy} shows
that saturation is reached most rapidly with
$\mathbf{H}_\mathrm{app}
\parallel [111]$, demonstrating that [111] is the ferromagnetic
easy axis at this temperature.

The principal cause of the rounding of the
$M_\parallel(H_\mathrm{int})$ curve, even for $H\,\parallel
[111]$, is likely to be the inhomogeneity of the demagnetizing
field. The only practical sample shapes that give a uniform
demagnetizing field are ellipsoids of revolution. We have
calculated the volume distribution of the demagnetizing field for
a perfect cylinder of the required aspect ratio assuming that the
magnetization is uniform; the characteristic width of the
distribution is $\delta D\sim0.07$ and, for example, $\sim12\%$ of
the sample volume has a local demagnetizing field that exceeds the
volume average value by a factor $>1.5$. Angular misalignments of
up to $2^\circ$ may also contribute.

The first two terms in the standard expansion of the anisotropic
free energy of a ferromagnet with cubic symmetry \cite{Williams37}
can be expressed, for the present geometry, as
\begin{equation}
F(\theta)=\sin^2\theta(\cos^2\theta+\frac{1}{4}\sin^2\theta)K_1+\frac{1}{4}(\sin^4\theta\cos^2\theta)
K_2
\end{equation}
where $\theta$ is the angle within the (110) plane between
$\mathbf{M}$ and [100] and $K_{1,2}$ are the first and second
anisotropy constants. It is possible, in principle, to determine
the anisotropy constants by assuming that $|\mathbf{M}|$ is fixed
in the `approach to saturation' region of the
$M_\parallel(H_\mathrm{int})$ curve \cite{Williams37}. In this
picture, the direction of $\mathbf{M}$ (and hence the magnitude of
$M_\parallel$) for any given $H_\mathrm{int}$ is determined from
the condition that the torque acting on $\mathbf{M}$ due to the
anisotropy exactly balances the torque due to the interaction of
$\mathbf{M}$ with $\mathbf{H}_\mathrm{int}$. However we choose an
alternative thermodynamic method \cite{Williams37} to extract
anisotropy constants from $M(H)$ which is less reliant on
assumptions about the demagnetizing field. We calculate the total
magnetic work done in order to bring the sample to saturation with
$\mathbf{H}_\mathrm{app}$ applied parallel to a certain crystal
direction. The work required to produce an infinitesimal increase,
$\mathrm{d}\mathbf{m}$, in the sample moment is $dW=\mu_0
\mathbf{H}_\mathrm{app}.\mathrm{d}\mathbf{m}$. Since the
irreversibility of $M(H)$ is negligible in ZrZn$_2$, we can relate
the magnetic work done to a change in the appropriate
thermodynamic function of state, namely the Gibbs free energy, and
we obtain $\Delta G=\int
\mu_0\mathbf{H}_\mathrm{app}\cdot\mathrm{d}\mathbf{m}$. The
anisotropy energy $F(\theta)$ is therefore given (apart from an
irrelevant constant) by the area under the
$H_\mathrm{app}(M_\parallel)$ curve. From Eq.~(2) it follows that
$K_1=4(F_{[110]}-F_{[001]})$ and
$K_2=9(F_{[001]}+3F_{[111]}-4F_{[110]})$. Evaluating the areas
corresponding to the magnetic work, we find
$F_{[110]}-F_{[111]}=1.1\times 10^{-8}$\,eV\,$\mathrm{f.u.}^{-1}$
and $F_{[100]}-F_{[111]}=1.6\times
10^{-8}$\,eV\,$\mathrm{f.u.}^{-1}$. The cubic anisotropy constants
are therefore $K_1=-2.0\times 10^{-8}$\,eV\,$\mathrm{f.u.}^{-1}$,
$K_2=-2.6\times 10^{-7}$\,eV\,$\mathrm{f.u.}^{-1}$. The magnitude
of the ferromagnetic anisotropy is therefore a factor $\sim 3$
smaller than that found in the weak itinerant ferromagnet Ni$_3$Al
[Ref.~\onlinecite{Sigfusson81}] (1\,erg cm$^{-3}\equiv
3.15\times10^{-11}$\,eV\,f.u.$^{-1}$ in ZrZn$_2$), where the
relative sizes of $K_1$ and $K_2$ are opposite to those in
ZrZn$_2$, and almost two orders of magnitude smaller than in Ni
[Ref.~\onlinecite{Franse68}].

The magnetocrystalline anisotropy of ferromagnetic metals can in
principle be obtained from band-structure calculations of the
total energy in which the effect of spin-orbit coupling is
included \cite{Brooks1940,Fletcher1954}. It would be interesting
to see whether such calculations correctly predict the magnetic
anisotropy of ZrZn$_2$.

\subsection{\label{sec:NS_res}Resistivity}
Many systems close to a quantum critical point have been shown to
exhibit a so-called non-Fermi liquid temperature dependence of the
resistivity.  Non-Fermi liquid is generally accepted to mean that
the low temperature exponent $n$ in the equation
$\rho(T)=\rho_{0}+A T^{n}$ is not equal to 2. Notable examples of
systems exhibiting non-Fermi liquid power laws in the resistivity
include high-temperature superconductors ($n \approx 1$ close to
optimal doping), heavy Fermion antiferromagnets, such as
CePd$_2$Si$_2$ [Ref.~\onlinecite{Mathur98}] ($n \approx 1.2$ at
high pressure close to quantum criticality), and the helical
ferromagnet MnSi [Ref.~\onlinecite{Pfleiderer97}]
($n\rightarrow\frac{5}{3}$ as $T\rightarrow 0$, in the
paramagnetic state close to the pressure-induced
ferromagnet-paramagnet QPT). The low moment ferromagnets Ni$_3$Al
and YNi$_3$ [$M(0,0)\approx0.075$ and
$0.04\,\mu_\mathrm{B}\,\mathrm{f.u.^{-1}}$ respectively] exhibit a
non-Fermi liquid resistivity over a wider range of pressures about
the QCP and, even at ambient pressure, a $T^{1.65}$ dependence is
observed over a small temperature range \cite{Steiner03}.

\begin{figure}
\begin{center}
\includegraphics[width=0.95\linewidth,clip]{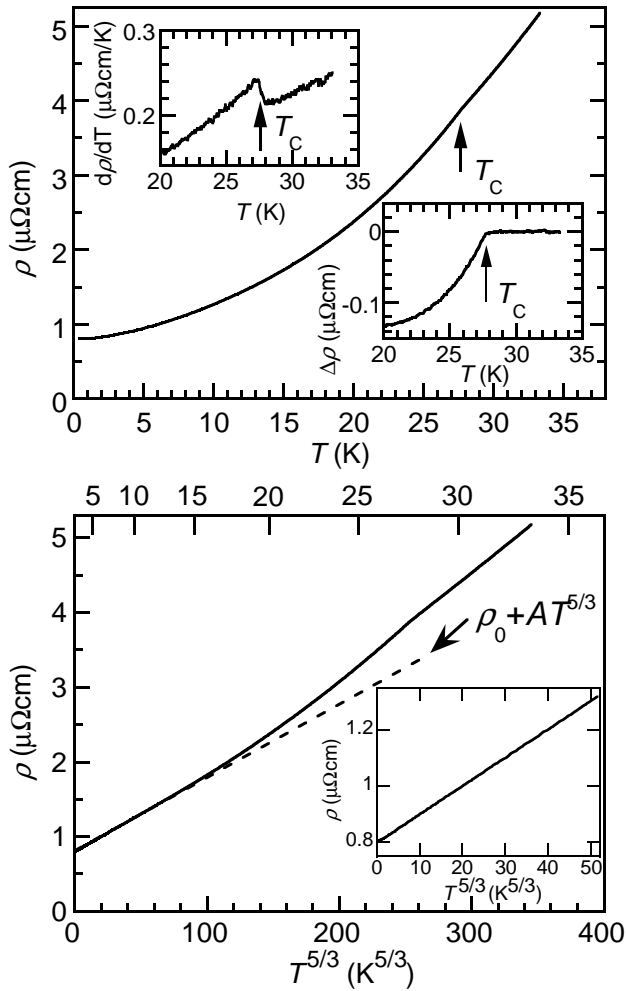}
\end{center}
\caption{ \label{fig:zrzn2_res_high_T} Upper panel: raw $\rho(T)$
results for $0.5<T<32$\,K; upper inset:
$\mathrm{d}\rho/\mathrm{d}T$ close to $T_C$; lower inset:
$\rho(T)$ results close to $T_C$ after subtraction of a smoothly
varying background (see text). Lower panel: $\rho(T)$ plotted
against $T^{5/3}$; inset: the same data over a limited $T$ range,
demonstrating the non-Fermi liquid power law
$\rho(T)=\rho_0+AT^{5/3}$.}
\end{figure}

Fig.~\ref{fig:zrzn2_res_high_T} shows the temperature dependence
of the resistivity for ZrZn$_2$.  On cooling the sample through
$T_C$, we observe a slight `kink' in the resistivity near the
Curie temperature. Similar behavior is observed in other itinerant
ferromagnets such as nickel, although the behavior in ZrZn$_2$ is
not as pronounced as that in stronger ferromagnets. The
resistivity anomaly associated with the ferromagnetic transition
can be emphasized if a background variation of the form $\rho_0 +
A T^2$ is subtracted from the data; the result is shown in the
lower inset. It has been argued that the dominant magnetic
contribution to the resistivity is due to short range spin
fluctuations and therefore $\mathrm{d}\rho/\mathrm{d}T$ should
vary like the magnetic specific heat near the critical point
\cite{Fisher68}. To test this hypothesis we have plotted
$\mathrm{d}\rho/\mathrm{d}T$ in the upper inset of
Fig.~\ref{fig:zrzn2_res_high_T}. There is good qualitative
agreement with the measured specific heat anomaly (see inset to
Fig.~\ref{f:zrzn2_C_T}).

At lower temperatures, in the ferromagnetic state, the resistivity
has a well defined non-Fermi liquid behavior. Fitting data over
the temperature range $1<T<14$\,K, we find $n=1.67\pm 0.02$. The
inset to Fig.~\ref{fig:zrzn2_res_high_T} shows that when $\rho$ is
plotted against $T^{5/3}$, a linear behavior is obtained up to
about 15\,K. The existence of a $T^{5/3}$ power law very close to
the ferromagnetic QCP has already been established in
high-pressure measurements on ZrZn$_2$
[Ref.~\onlinecite{Grosche95}] but our result shows that the
unusual power law applies well away from the QCP. Early data on a
polycrystalline sample \cite{Ogawa76} showed a $T^{5/3}$
temperature dependence in the range $15<T<50$\,K but notably not
at lower temperatures.

It has been pointed out by a number of workers that a $\rho
\propto T^{5/3}$ behavior may be understood in terms of scattering
of quasiparticles by spin-fluctuations \cite{Mathon68}. A
$T^{5/3}$ dependence is expected near a 3D ferromagnetic quantum
critical point where the collective spin excitations are expected
to be overdamped.  Thus it appears that ZrZn$_2$ is sufficiently
close to quantum criticality to observe the 5/3 non-Fermi liquid
behavior.

It has been reported that ZrZn$_2$ displays superconductivity
below about 0.6\,K. We find no traces of superconductivity in
our resistivity curves at
ambient pressure in samples etched in a HF/HNO$_3$ solution
\cite{Yelland05}.

\subsection{\label{sec:spec_heat} Heat Capacity}
\begin{figure}
\begin{center}
\includegraphics[width=0.95\linewidth,clip]{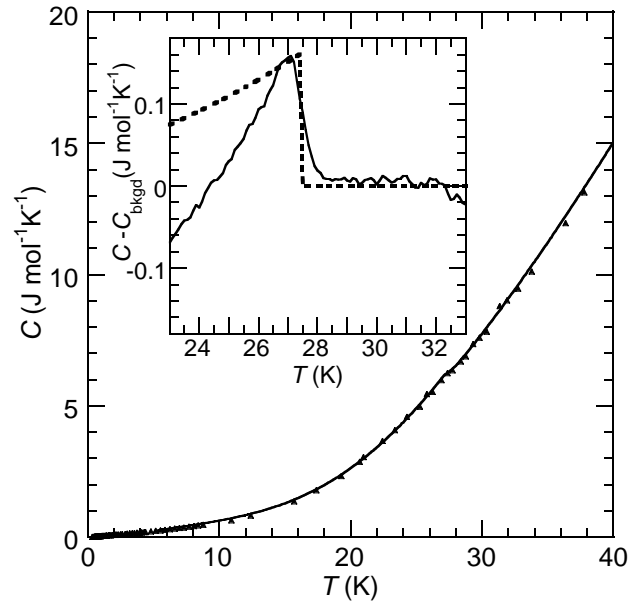}
\end{center}
\caption{ \label{f:zrzn2_C_T} Specific heat $C(T)$ of single
crystal ZrZn$_2$: symbols show data obtained using a
pulse-relaxation technique; solid line shows the results obtained
by a high sensitivity a.c.\ method. Inset: solid line shows a.c.\
data close to $T_C$ after subtracting a smoothly varying
background (see text); dashed line shows $C_\mathrm{m}(T)$
calculated by Stoner theory, multiplied by a scale factor 0.5.}
\end{figure}

Fig.~\ref{f:zrzn2_C_T} shows specific heat results for $T<40$\,K
on single crystal samples that were cut from a region of the ingot
next to that used for resistivity measurements. As with previous
measurements \cite{Pfleiderer01c} a large linear contribution to
the specific heat $C=\gamma T$ was observed at low temperatures,
with $\gamma=$\ 45\,mJ\,K$^{-2}$\,mol$^{-1}$.  At higher
temperatures the high-sensitivity of a.c.\ specific heat
measurements allows the small ferromagnetic anomaly to be
observed.

The anomaly in $C(T)$ at the ferromagnetic transition is only
$2.5$\% of the total in ZrZn$_2$; a reliable separation of the
magnetic heat capacity $C_\mathrm{m}$ therefore means that both
the electronic and phonon components must be determined precisely,
and this is not a trivial task. In order to emphasize the
ferromagnetic anomaly, in the inset to Fig.~\ref{f:zrzn2_C_T} we
have plotted the heat capacity after subtracting a smoothly
varying estimate of the non-magnetic heat capacity of the form
$C_\mathrm{bkgd}(T)=\gamma T + C_\mathrm{Debye}(T/\Theta)$; here
$C_\mathrm{Debye}$ is the Debye heat capacity function and
$\Theta$ is a Debye temperature. We set
$\gamma=38$\,mJ\,K$^{-2}$\,mol$^{-1}$, close to that observed at
low temperatures, in order to obtain a flat $C_\mathrm{m}(T)$
immediately above $T_C$; the only other free parameter $\Theta$
was allowed to vary to give the best fit to the data in a small
temperature range $29<T<33$\,K above $T_C$, giving a value
$\Theta=284$\,K. We emphasize that this estimate of the
non-magnetic heat capacity may not be correct in detail (e.g.
there may be significant fluctuation heat capacity in the range of
the fit and we have not taken into account the true phonon DOS),
but our quantitative discussion will be confined to the height of
the discontinuity at $T_C$ which is hardly affected by the choice
of background.

The jump in $C$ at $T_C$ measured here is about a factor of 2
larger than that measured previously \cite{Viswanathan74} on a
sample with $T_C=10$\,K. The shape of the anomaly, shown in the
inset to Fig.~\ref{f:zrzn2_C_T}, broadly resembles that expected
for a mean-field second order transition. However, closer
inspection shows that there is curvature in our estimate of
$C_\mathrm{m}(T)$, both above and below the jump at $T_C$, up to
1.5\,K from $T_C$.  This almost certainly results from thermal
fluctuations, although our uncertainty in the background precludes
a detailed analysis. There is also evidence of some rounding of
the anomaly from sample inhomogeneity on a smaller temperature
scale $\sim0.2$\,K.

We now compare our results with Landau and spin fluctuation
theories. In the Landau approach, the free energy is written in the
form
\begin{eqnarray}
\label{dis:landau_F} F & = & \frac{1}{2} a(T) M^2 + \frac{1}{4} b
M^4 - M B.
\end{eqnarray}
First, we relate the parameters of the theory, namely the Landau
coefficients $a(T)$ and $b$, to experimental quantities determined
near $T_C$. In mean field theory one usually assumes that the
coefficient $a(T)$ varies linearly with temperature near $T_C$ so
that $a(T)=\dot{a} t$, where $t=(T-T_C)/T_C$. By minimizing the
Landau free energy Eq.~\ref{dis:landau_F} with $B=0$, we find the
spontaneous magnetization is
\begin{eqnarray}
\label{dis:landau_M} M & = & \sqrt{-\frac{\dot{a} t}{b}}.
\end{eqnarray}
Combining Eq.~\ref{dis:landau_F} and Eq.~\ref{dis:landau_M}, the
zero-field specific heat can be evaluated from the thermodynamic
relation $C_V=-T(d^2F/dT^2)$,
\begin{eqnarray}
\label{dis:landau_C}
C_\mathrm{m}^\mathrm{(Landau)} & = & \frac{\dot{a}^2 T}{2b T_C^2} \ \ \mathrm{for} \: T < T_C, \nonumber \\
  & = & 0 \ \ \ \ \ \ \ \,\mathrm{for} \: T > T_C.
\end{eqnarray}
Since this simple model only includes terms in the free energy
that are dependent on the macroscopically ordered moment,
$C_\mathrm{m}$ must vanish above $T_C$; the discontinuity in the
specific heat at $T_C$ is simply the limiting value as
$T\rightarrow T_C$, i.e.\ $\Delta C_\mathrm{m}=\dot{a}^2/2b T_C$.
This value can easily be compared with experiment by noting that
the gradient of the critical isotherm in the Arrott plot
Fig.~\ref{fig:zrzn2_arrott} gives $1/b$ and that in the
magnetization scaling plot Fig.~\ref{fig:zrzn2_scaling}, the
intercept of the $T<T_C$ data with the ordinate axis gives
$\sqrt{\dot{a}/b}$ directly. From the data we obtain $b=1.1\times
10^{-13}$\,T\,(Am$^{-1}$)$^{-3}$ and $\sqrt{\dot{a}/b}=4.0\times
10^4$\,Am$^{-1}$. This gives a discontinuity $\Delta
C_\mathrm{m}^\mathrm{(Landau)}=150$\,mJ\,K$^{-1}$\,mol$^{-1}$, in
excellent agreement with the experimental value of 155 $\pm$
30\,mJ\,K$^{-1}$\,mol$^{-1}$.  The agreement between our measured
$\Delta C$ and that calculated from the magnetization isotherms
provides a consistency check on the Landau theory.

The specific heat anomaly in weak itinerant ferromagnets has
been the subject of various theoretical studies \cite{Wohlfarth77,
Makoshi75,Mohn89,Ishigaki99,Ishigaki96,Takahashi04}.
In the Stoner-Wohlfarth model \cite{Wohlfarth77}, the Landau coefficient $a$
has the temperature dependence


\begin{eqnarray}
\label{dis:eq2} a(T) & = & -\frac{\mu_0}{2 \chi_0} \left(
1-\frac{T^2}{T_C^2} \right),
\end{eqnarray}
In this model the exchange field $\Delta$ is
proportional to the magnetization; the $T$ dependence of $a$
reflects the reduction in magnetization due to thermal
spin-flipping excitations. The $b$ coefficient can be obtained by
minimizing $F$ at $T=0$, giving
\begin{eqnarray}
\label{dis:eq3} b & = & \frac{\mu_0}{2 \chi_0 M_0^2},
\end{eqnarray}
where
\begin{eqnarray}
\label{dis:eq4}
\chi_0 & = &\left( \frac{dM}{dB} \right)_{B,T \rightarrow 0},
\end{eqnarray}
and $M_0$ is the $T=0$ magnetization.  As in the Landau model, we
can estimate the specific heat from $F(T)$
[Ref.~\onlinecite{Wohlfarth77}],
\begin{eqnarray}
\label{dis:Stoner_C} C_\mathrm{m}^\mathrm{(Stoner)} & =
&-\frac{M_0^2}{2 \chi_{0} T_C}
 \left( \frac{T}{T_C} - 3 \frac{T^3}{T_C^3} \right) \mathrm{for} \: T < T_C, \nonumber \\
  & = & 0 \ \ \ \ \ \ \ \ \ \ \ \ \ \ \ \ \ \ \ \ \ \ \ \ \ \ \ \:\:\:\:\mathrm{for}\ T > T_C.
\end{eqnarray}

Using $M_0=3.1\times 10^4$\,Am$^{-1}$
($\equiv0.17\,\mu_\mathrm{B}$\,f.u.$^{-1}$) and $\chi_0=4.2\times
10^{-3}$, as determined from $M(H)$ measurements at $T=1.8$\,K
[Ref.~\onlinecite{Pfleiderer01a}], we find $\Delta
C_\mathrm{m}^\mathrm{(Stoner)}=330$\,mJ\,K$^{-1}$\,mol$^{-1}$
which is a factor 2 larger than the experimental value.

The Stoner-Wohlfarth approach can only include the effect of spin
fluctuations through the renormalization of its phenomenological
parameters. Experiments have shown that many of the properties of
weakly ferromagnetic materials such as ZrZn$_2$ and Ni$_3$Al
cannot be explained within this framework.  Perhaps the most
obvious property not explained by a mean field approach is the
temperature dependence of the susceptibility above $T_C$, for
which experiment shows a Curie-Weiss dependence $\chi \propto
(T-T_C)^{-1}$ as opposed to the $\chi \propto (T-T_C)^{-2}$
predicted by Stoner theory. In order to address the deficiencies
of mean field theories, various self-consistent renormalized (SCR)
spin fluctuation models have been proposed
\cite{Murata72,Lonzarich85,Moriya1985Book}. These follow the
Landau-Ginzburg approach and treat the local magnetization
$\mathbf{m(r)}$ as a fluctuating stochastic variable. The SCR
theory allows both magnetic corrections above $T_C$ and the
renormalizing effect of spin fluctuations on the Landau $a$
parameter to be taken into account. The effect of including spin
fluctuations is to reduce the discontinuity in the specific heat
$\Delta C_\mathrm{m}$ at $T_C$.  This has been estimated by Mohn
and Hilscher \cite{Mohn89} based on the model of Murata and
Doniach \cite{Murata72},
\begin{equation}
\label{dis:SCR_C} \Delta C_\mathrm{m}^\mathrm{(SCR)}  =
-\frac{\mu_0 M_0^2}{4 \chi_{0} T_C}.
\end{equation}
Note that the Mohn and Hilscher value $\Delta
C_\mathrm{m}^\mathrm{(SCR)}$ =(1/4) $\Delta
C_\mathrm{m}^\mathrm{(Stoner)}$. The experimental value of $\Delta
C_\mathrm{m}$ and the various model predictions are summarized in
Table~\ref{table:C}.
\begin{table}
\caption{\label{table:C} Comparison of calculated and measured
specific heat discontinuities.}
\begin{ruledtabular}
\begin{tabular}{c|cccc}
  & Landau  &  Stoner & SCR & Exp  \\
\colrule
$\Delta C( \mathrm{mJ\,K^{-1}\,mol^{-1}}$) & 0.15  & 0.33 &  0.081 & 0.155 \\
\end{tabular}
\end{ruledtabular}
\end{table}

In summary, the Stoner-Wohlfarth model overestimates the specific
heat jump and the SCR-theory, as implemented in
Refs.~\cite{Murata72,Mohn89}, underestimates the jump.  It is
pleasing to note however that the shape of the anomaly predicted
by Stoner theory is similar to that observed experimentally.
Because of the nature of the theories one cannot attribute their
failure to a single approximation.  In the case of the Stoner
theory the phenomenological parameters ($T_C$, $\chi_0$, $b$) used
as inputs to the model are not taken directly from band structure
calculations, rather they are determined from the experimental
magnetization $M(B,T)$ and thus can be renormalized by
fluctuations. The fact that the Stoner-Wohlfarth theory
overestimates the specific heat jump suggests that paramagnetic
correlations give a significant contribution to the specific heat
above $T_C$. The SCR-theory might be improved by using a more
realistic model for the excitations
($\chi^{\prime\prime}(q,\omega)$) below $T_C$, where the
longitudinal and transverse excitations are treated separately.
Unfortunately, a full implementation of the SCR theory including,
for example, the effects of the changes in the magnetic excitation
spectrum on entering the ferromagnetic state is difficult
\cite{Lonzarich85}.

\subsection{Conclusion}

In conclusion we have studied various thermodynamic and transport
properties of the weak itinerant ferromagnet ZrZn$_2$.
Magnetization measurements show that the easy crystallographic
axis is [111] and scaling plots reveal mean-field exponents for
the temperature range studied.  Specific heat measurements reveal
an anomaly whose shape is reminiscent of mean-field behavior.
However, the measured discontinuity is significantly smaller than
that predicted by Stoner-Wohlfarth theory. The resistivity shows
an anomaly at $T_C$ and a non-Fermi liquid $T^{5/3}$ behavior at
low temperatures.  Our results demonstrate the importance of
collective spin fluctuations in ZrZn$_2$, a material that was once
considered to be a candidate for a Stoner ferromagnet.

\subsection*{ Acknowledgments}

We wish to thank N.R. Bernhoeft, G.G. Lonzarich, P.J. Meeson, and
C. Pfleiderer for informative discussions and help with this work,
and C. Pitrou for performing some preliminary heat capacity
experiments. The research has been supported by the EPSRC.


\begin{thebibliography}{58}
\bibitem[{dHv()}]{dHvAmassnote}
\bibinfo{note}{At a finite field $B\approx 9$\,T, de Haas-van Alphen
  measurements give quasiparticle masses that are enhanced above band-structure
  values by a factor 3--5, in good agreement with heat capacity measurements at
  this field \cite{Pfleiderer01c} which suggest $1+\lambda \approx 3.7$}.
\end{thebibliography}
\end{document}